%Paper: astro-ph/9308026
%From: davet@granta.uchicago.edu (David Thomas)
%Date: Fri, 20 Aug 93 12:16:05 CDT

% Production of Li, Be \& B from Baryon Inhomogeneous
% Primordial Nucleosynthesis
% Plain TeX

% @(#)paper.tex	1.2 8/4/93

% Page size
\magnification=\magstep1
\vsize9truein
\hsize6.5truein
% \abovedisplayskip=18pt plus 5pt minus 13pt  % Leave more space around
% \belowdisplayskip=18pt plus 4pt minus 13pt  %     displayed equations

% Fonts
                        % for use in text subscripts
                     % for notes
% \font\titlefont=cmbx12 at 18truept
% \font\truetenrm=cmr10 at 10truept
% \font\truetenbf=cmbx10 at 10truept
\def\doublespacing{\baselineskip=20pt}
\def\singlespacing{\baselineskip=12pt}

% Elements
\def\h#1{${}^{#1}$H}
\def\D{D}
\def\he#1{${}^{#1}$He}
\def\li#1{${}^{#1}$Li}
\def\be#1{${}^{#1}$Be}
\def\b#1{${}^{#1}$B}
\def\c#1{${}^{#1}$C}
\def\n#1{${}^{#1}$N}
\def\o#1{${}^{#1}$O}
\def\f#1{${}^{#1}$F}
\def\ne#1{${}^{#1}$Ne}

% Units

% Misc
\def\refitem{\par\noindent\hangindent=30pt\hangafter=1}
\def\section#1{{\bigskip\bigskip\centerline{\bf #1} \bigskip}}
\hyphenation{steig-man}
\def\ref{{\it ref}}

% lsim and gsim are borrowed from TeX by Example
\def\stacksymbols #1#2#3#4{\def\theguybelow{#2}
    \def\verticalposition{\lower#3pt}
    \def\spacingwithinsymbol{\baselineskip0pt\lineskip#4pt}
    \mathrel{\mathpalette\intermediary#1}}
\def\intermediary#1#2{\verticalposition\vbox{\spacingwithinsymbol
      \everycr={}\tabskip0pt
      \halign{$\mathsurround0pt#1\hfil##\hfil$\crcr#2\crcr
               \theguybelow\crcr}}}
\def\lsim{\stacksymbols{<}{\sim}{2.5}{.2}}
\def\gsim{\stacksymbols{>}{\sim}{3}{.5}}

%------------------ Title Page -------------------------------------------

\singlespacing
\rightline{UMN-TH-1131/93}
\rightline{astro-ph/9308026}
\rightline{JULY 1993}
\doublespacing

\vskip 1cm
\centerline{\bf Production of Li, Be \& B from }
\centerline{\bf Baryon Inhomogeneous Primordial Nucleosynthesis}

\vskip1cm
\centerline{David Thomas${}^1$, David N.~Schramm${}^{1,2}$,
            Keith A.~Olive${}^3$,}
\centerline{Grant J.~Mathews${}^4$, Bradley S.~Meyer${}^5$,
            Brian D.~Fields${}^1$}

\vskip1cm
\singlespacing
\centerline{${}^1$ University of Chicago, Chicago, IL 60637}
\centerline{${}^2$ NASA/Fermilab Astrophysics Center, FNAL, Box 500
            Batavia, IL 60510}
\centerline{${}^3$ University of Minnesota,
            School of Physics and Astronomy, Minneapolis, MN 55455}
\centerline{${}^4$ Lawrence Livermore National Laboratory, Livermore, CA 94550}
\centerline{${}^5$ Department of Physics and Astronomy, Clemson University,
            Clemson, SC 29634-1911}

\section{Abstract}

\midinsert\narrower
We investigate the possibility that inhomogeneous nucleosynthesis may
eventually be used to
explain the abundances of \li6, \be9 and B in
population II stars.  The present work differs from previous studies in that
we have used a more extensive reaction network.
It is demonstrated that in the simplest scenario
the abundances of the light elements with $A\le7$ constrain the separation of
inhomogeneities to sufficiently small scales that the model is
indistinguishable from homogeneous nucleosynthesis and that the abundances
of \li6, \be9\ and B are then below observations by several orders of
magnitude.  This conclusion does not depend on the \li7 constraint.
We also examine
alternative scenarios which involve a post-nucleosynthesis reprocessing
of the light elements to reproduce the observed abundances of Li and B,
while allowing for a somewhat higher baryon density (still well below
the cosmological critical density).  Future B/H measurements
may be able to exclude even this exotic scenario and further restrict
primordial nucleosynthesis to approach the homogeneous model conclusions.
\endinsert

\bigskip
\hrule width 3truein
\smallskip
\singlespacing
\leftline{Submitted to {\it The Astrophysical Journal}}
\doublespacing

\vfil\eject

%------------------ First Page -------------------------------------------

\section{1. Introduction}
There has been considerable recent interest in the possibility that baryons
may have been distributed inhomogeneously  in the early universe.  There are
a number of mechanisms by which such inhomogeneities could be produced
(c.f.\ Malaney \& Mathews, 1993).  Perhaps the most frequently considered
has been a first-order QCD phase transition.  It is
quite possible that density inhomogeneities could be produced (Crawford and
Schramm 1982; Hogan 1983; Witten 1984; Iso, Kodama \& Sato 1986; Fuller,
Mathews \& Alcock 1988; Kurki-Suonio 1988; Kapusta and Olive 1988).
These perturbations may have had a
profound effect on the production of the light elements
in the early universe (Applegate and Hogan 1985; Sale and Mathews
1986; Applegate, Hogan \& Scherrer 1987; Alcock, Fuller \& Mathews
1987; Malaney and Fowler 1988; Kurki-Suonio and Matzner 1989 and 1990;
Terasawa and Sato 1989a,b,c; Kurki-Suonio, Matzner, Olive \& Schramm
1990; Mathews, Meyer, Alcock \& Fuller 1990).  In this paper we examine
the consequences of inhomogeneous nucleosynthesis on the intermeditate
stable isotopes \li6, \be9, \b{10} and \b{11}.

In recent years measurements
have been made in population II stars
of elements that were once thought to have been produced in
insignificant quantities in the standard homogeneous big bang.
One of these, \be9 (Rebolo et al.\ 1988; Ryan et al.\ 1992;
Gilmore, Edvardsson, \& Nissen 1991)
has been proposed (Boyd, \& Kajino 1989; Malaney, \& Fowler 1989) as a
potential signature of baryon-inhomogeneous nucleosynthesis.
However the strong observed correlation
of Be/H with metallicity implies that it was made in the early Galaxy
(Walker et al.\ 1993, Fields, Schramm \& Truran 1993).
To date, population II B/H data includes
only three measurements (Duncan, Lambert, \& Lemke 1992), which again
show a correlation with metallicity.  There has been one observation
so far of \li6 (Smith, Lambert, \& Nissen 1992).

These data appear to be best explained by galactic cosmic ray
spallation (Steigman \& Walker 1992; Walker et al.\ 1993; Prantzos,
Cass\'e, \& Vangioni-Flam 1993; Olive and Schramm 1992; Steigman et al. 1993),
however there remains the question of whether \li6, Be and B can be produced by
primordial nucleosynthesis, and if so whether they can provide the
much sought after litmus test of baryon inhomogeneities in the early
universe.  Four of us (Thomas et al.\ 1993, hereafter TSOF) used the largest
network to date (in terms of reactions influencing light, $A\le12$,
element abundances)
to demonstrate that standard homogeneous big bang
nucleosynthesis underproduces these nuclei by at least 2 (\li6), 4
(\be9) and 5 (B) orders of magnitude, when the abundances of the
lighter elements and \li7 are compared to the observations.
In this paper we investigate
inhomogeneous yields using an even further extension of
the reaction network developed in TSOF.
Our work primarily differs from previous similar
studies (e.g.\ Terasawa \& Sato, 1989c)
in that we have considered a larger network of reactions to produce
\li6, \be9, \b{10} and \b{11}, fully allowing for neutron-rich flows
and multiple back-reactions.

Nucleosynthesis in a homogeneous big bang requires the evolution of a
set of equations representing the rates of the nuclear reactions in
the network.  The only input parameter is the baryon to photon
ratio $\eta=n_B/n_\gamma$ or equivalently the density of baryons,
(since $n_\gamma$ can be directly related to the microwave background
temperature).
Additional parameters are introduced when the effects of baryon
inhomogeneities are taken into account.
One of these parameters is the
length scale, $l$, associated with the fluctuations. If this length scale
is much greater than the neutron diffusion length then diffusion
is unimportant and the yields can be obtained by simply averaging the yields
from regions with different baryon densities (Wagoner, 1973; Yang et
al. 1984).  In the homogeneous
model however, data on D and He restrict the baryon density to
a small range about $\eta_{10}=3$ ($\eta_{10} = \eta/10^{-10}$) where
\li7 takes on its minimum value ($\hbox{\li7}/\hbox{H}\sim 10^{-10}$)
in agreement with observations (Spite \&
Spite 1982a,b,1986; Hobbs and Duncan 1987; Rebolo, Molaro \& Beckman 1988).
 Any average of the $\eta_{10}=3$
solution with that for another value of $\eta$ will increase the \li7
abundance, and is likely to violate the upper limit ($1.4\times10^{-10}$).
Thus one obtains strong constraints on the amplitude of such perturbations.

In the other extreme (inhomogeneities much smaller than the neutron
diffusion length) baryon
diffusion will eradicate the inhomogeneities before nucleosynthesis begins,
and we return to the homogeneous case.
It is the intermediate
case that interests us here.  When the inhomogeneity scale is of the
same order as the neutron diffusion length the more rapid diffusion of neutrons
(compared to protons) leads to an inhomogeneity in $n/p$ in addition
to the density inhomogeneity.  The earliest studies of this scenario
(Applegate, Hogan \& Scherrer 1987; Alcock et al.\ 1987)
assumed that neutrons diffused to a homogeneous density before nucleosynthesis
began, and neglected all diffusion effects during nucleosynthesis.
This simple model was able to satisfy constraints from \h2 and \he4
{(and \he3)} with a baryon density equal to the critical value
($\Omega_B=1$) and a density contrast ($\sim 100$) that seemed not
implausible from the point of view of quark-hadron physics.
This scenario was attractive because it did away with the need for
non-baryonic dark matter.  Unfortunately, it overproduced
\li7.  It was later suggested that the excess Li may be removed
by diffusion of neutrons back into the high density region after
nucleosynthesis (Malaney and Fowler 1988) for values of $l \sim 10$ m.
Though this turns out to
be only partially true (Terasawa \& Sato, 1989; Kurki-Suonio \& Matzner, 1989;
Mathews et al., 1990),
the important lesson is that an accurate
determination of
abundances requires a calculation which takes careful account of the diffusion
of neutrons and protons before, during and after nucleosynthesis.

In recent years, more detailed diffusion calculations (Kurki-Suonio
and Matzner 1989 and 1990; Terasawa and Sato 1989a,b,c; Mathews,
Meyer, Alcock \& Fuller 1990; Kurki-Suonio, Matzner, Olive, \& Schramm,
1990)
have shown that not only could \li7 be affected
but \he4 as well. It was found that
nucleosynthesis with $\Omega_B = 1$, no matter what the density contrast,
overproduced both \he4 and
\li7.\footnote{$^\dagger$}{The problem with \he4 is particularly important
since it does not allow for the possibility that consistency of all
the light elements is achievable simply by the depletion of \li7 in
non-standard solar models}
Indeed these latest calculations all showed that for $\Omega_B = 1$,
and when the distance scale of the  inhomogeneities,  $l$ is greater
than 30 m
only the D abundance can be brought into agreement  with  observations.
Though the
standard model constraints on $\eta$ can be modified, the modification was
shown to be rather limited (Kurki-Suonio, Matzner, Olive \& Schramm 1990).

In this paper we have used the diffusion code developed by Mathews et
al.\ (1990).
Initial density fluctuations are arranged in a lattice of spheres with
separation $l=2r$.  Each sphere is described as a high density core
and a low density outer shell.  The core has density and radius of
$R$ and $f_v^{1/3}$ respectively, relative to the outer region.
The sphere is divided up into concentric spherical zones, with a higher
resolution near the boundary between high and low density regions. We expect
the choice of spherical boundary conditions to
maximize the potential effect.
In all results presented here we have used 16 concentric zones.  We have
also run test cases with 8 and with 32 zones, indicating that 16 provide
adequate accuracy, while remaining economical with computer time
(Mathews et al.\ 1990). We
also note that results of this code are consistent with those of
Terasawa and Sato (1989a--c) and
Kurki-Suonio and Matzner (1989).

To alleviate some of the problems encountered in inhomogeneous
models, mechanisms have been proposed to reprocess the nucleosynthesis
products subsequent
to the epoch of primordial nucleosynthesis.
These mechanisms in particular reduce the abundance of \li7,
thus (potentially) allowing for higher $\Omega_B$ models.
One such mechanism (Alcock et al., 1990; Jedamzik, Fuller \& Mathews, 1993a)
examines a
fluid mechanical property of the electromagnetic plasma
near the end of nucleosynthesis.
The photon mean free path $\lambda_\gamma$
and the average physical (i.e.\ not comoving) size $l_h$ of the
high density regions have different temperature dependences.  At high
temperatures ($T \gsim 20$ keV), $\lambda_\gamma < l_h$, and so the
EM plasma is confined over regions smaller than the baryon fluctuations,
thus preserving these fluctuations.  Below
$T = T_m \sim 20$ keV, however,
$\lambda_\gamma > l_h$, and the EM plasma is not confined on the fluctuation
scales.  Protons in the high density regions flow out, hindered only
by radiation (Thomson) drag.   Alcock et al.\ (1990) argue that the
dissipation of the fluctuations will homogenize the universe.  They model
this effect by running the inhomogeneous code to a given $T_m$,
then following the
rest of the evolution in the standard (homogeneous) code.
They find that for the
favored range of $T_m$ there is a significant reduction in the final
abundances of Li, Be, and B over inhomogeneous production without dissipation,
with \li7/H in particular reduced from $\sim 10^{-9}$ to $\sim 10^{-10}$.
However, detailed calculations of Jedamzik et al.\ (1993) have shown that
this mechanism is not as efficient as was previously believed.

Gnedin and Ostriker (1993) have suggested another model of reprocessing,
in which a baryon rich universe ($\Omega_B^0 \simeq 0.15$) overproduces
\he4 and \li7, while underproducing \h2 and \he3.  They then posit that
Jeans-mass black holes are formed at recombination.  The black holes form
accretion disks which emit a photon flux and reprocesses the ambient
material; in particular, photodissociating the light elements and
producing \h2 and \he3 by dissociating \he4.  The net effect could be to
reproduce the observed levels of \h2 and \he3, while still overproducing
\li7 by a factor of 10, and producing \he4 at a level of $Y_p\simeq0.250$.

We will discuss the implications of reprocessing on our conclusions.

\section{2. The calculation}
Based on the reaction network developed earlier (TSOF) and extended where
necessary to allow for neutron-rich flows etc., we have
evaluated the yields of the light elements from inhomogeneous
primordial nucleosynthesis.  The earlier reaction network contained
180 reactions, see table 1 of TSOF.  However, as pointed out
in TSOF, the ``flow'' to
the heavier elements (Be, B) lies largely along the neutron-rich side
of the network, and thus mainly occurs in the low density, neutron rich
zones of the inhomogeneous model.
We felt it wise therefore to update the network
further.  The 84 additional reaction rates were estimated using
the methods outlined in TSOF and Fowler and Hoyle (1964), and are
shown in table 1.  The full network is shown in figure 1.  We have run a
few sample cases without the extra reactions, and find no significant
effect on the results.

The diffusion code of Mathews et al.\ (1990)
includes full multi-zoning and
neutron back-diffusion.  The diffusion coefficients used were those
calculated by Banerjee \& Chitre (1991), and Kurki-Suonio et al.\ %
(1992).  Results were obtained for a wide range of values for $\eta$
and $r$ with the density contrast fixed at
$R=10^2,10^3$, and fractional volumes (for the high density
region) $f_v=1/64,1/8$.  We have also calculated results for $R=10^6$ with
$f_v=1/64$.
The geometry assumed was spherical, with 16
concentric zones and a high density core.  Due to the small uncertainty
in the neutron mean life ($\pm$ 2.1 sec) we fix this value at 889.1
s. (Particle Data Group, 1992).

Baryon inhomogeneities have been best motivated by a possibility of a
first-order
QCD phase transition. Though the values of $R$ and $r$ can not yet be reliably
predicted by QCD,
some estimates can be made. For example, if chemical equilibrium is maintained,
the value
of $R$, which is very sensitive to the transition temperature $T_c$,is found
to be
between $7 < R < 100$ for
$T > 100$ MeV (Alcock et al., 1987; Kapusta \& Olive, 1988). More generally,
the value of $R$ is determined by a combination of the enhanced
thermodynamic solubility of baryon number in the high-temperature phase
and the limited baryon number permeability of the moving phase boundary.
Depending on the efficiency of baryon transport and the baryon
penetrability of the phase boundary, $R$ may be considerably larger (Witten,
1984; Fuller et al., 1988; Kurki-Suonio, 1988).  The ultimate value at
the time of nucleosynthesis, however, is expected to be less than $10^6$
due to the
effects of neutrino-induced heating and expansion of the fluctuations
(Heckler \& Hogan, 1993; Jedamzik et al., 1993a,b). The baryon number build-up
at the boundary surface (where $R$ is largest) contains only a small
fraction of the total baryon number (Kurki-Suonio 1988). Thus, though we
include
values of $R$ as large as $10^6$ in our calculations, this should be viewed
as an extreme
upper limit.

The value of $r$ is also very sensitive to $T_c$ and the surface
tension, $\sigma$, of the phase interface (Fuller, Mathews \& Alcock 1988;
Kajantie, Karkkainen \& Rummukainen 1990); $r  \simeq 2 \times 10^4
m ({\sigma \over MeV^3})^{3/2} ({T_c \over MeV})^{-13/2}$.
For values of $\sigma^{1/3} \simeq  70$ MeV estimated
by Fahri and Jaffe (1984) which agree with the effective field theory model
estimates (Campbell, Ellis \& Olive 1990),
$r  \lsim 0.4 m$ for $T_c \gsim 100$ MeV.  This is to be compared with
preferred values of $r \approx 30$ m or the more recent estimates
(Kurki-Suonio et al. 1992) of $r  \approx 100$ m
at which reductions (though still insufficient)
in the production of \he4 and \li7 occur.
It is important to note that the available estimates from QCD are all
perfectly compatible with {\it homogeneous} nucleosynthesis.

\section{3. Results}
Results are shown in figures 2--8.  Figure 2a shows the $\eta$--$r$ plane
(where $\eta$ is the baryon to photon ratio, $n_B/n_\gamma$ and $r$ is
the radius of the spherical regions in cm,
measured at 100 MeV, after the phase transition) for $R=100$
(results for $f_v=1/8$ and $f_v=1/64$ are combined).  The contours show
observational limits on the abundances of the light elements
(Walker et al. 1991 (WSSOK) and refs.\ therein):
$$0.22\le Y_p\le0.24\eqno{(1)}$$
$$^2\hbox{H}/\hbox{H}\ge1.8\times10^{-5}\eqno{(2)}$$
$$(^2\hbox{H}+^3\hbox{He})/\hbox{H}\le1.0\times10^{-4}\eqno{(3)}$$
$$1.0\times10^{-10}\le^7\hbox{Li}/\hbox{H}\le1.4\times10^{-10}.\eqno{(4)}$$
In addition, the dashed curve represents a He mass fraction $Y_p=0.245$
which is the most recently
derived (preliminary) upper limit on $Y_p$ (Skillman et al. 1993).  The region
which
satisfies all these constraints is hatched.  Note that the only effect of
increasing the maximum He abundance to 0.245 is to allow a slightly
higher value of $r$.  Figure 2b shows similar data for $R=1000$,
and 2c for $R=10^6$.
Note that in figures 2 the hatched regions cover a similar area.

For small $r$ ($\lsim100$ cm), diffusion eliminates inhomogeneities before
nucleosynthesis begins and the results are identical to those from a
homogeneous calculation.  As $r$ increases from 100 cm, the He mass fraction
rises rapidly above 0.24.  Since all curves in figures 2 are parallel
to the $r$ axis for $r\lsim 100$ cm,
we conclude that any inhomogeneous model that satisfies
the limits on light element abundances will give the same abundances as
the homogeneous model, and the same limits on $\eta$
($2.8\lsim\eta_{10}\lsim3.3$, WSSOK).  Since the hatched regions
cover an almost identical area, this conclusion is independent of
$R$ and $f_v$.  We have also verified that in the large $r$ limit (no
diffusion) yields become independent of $r$.

If we relax the upper limit on \li7, (say, because of some subsequent
Li destruction) there is little change unless
we also relax the upper limit on \he4.  The dashed curve in figures 2 represent
a \he4 mass fraction of 0.245.  In the case where $Y_p < 0.245$ there are
two allowed
regions if we allow the primordial abundance of \li7
to exceed $4 \times 10^{-10}$:
(1) the previous limits are now $2.8\lsim\eta_{10}\lsim6$
and $r\lsim100$ cm;
(2) there is a region between the \h2+\he3 curve and the dashed
curve at $\eta\sim7$, $r\sim10^4$.  This solution however requires a rather
finely tuned value of $r$ in addition to the excess production of \li7, which
would require the depletion of \li7 by more than factor of 4
(we note that standard stellar models (Deliyannis, Demarque \& Kawaler
1990) do not deplete \li7 significantly and
non-standard stellar models which do deplete \li7 are highly constrained
by the observation of \li6 in HD 84937 (Steigman et al. 1993)).
Furthermore, since the code calculates abundances for a
uniform lattice of spheres an accurate determination of yields
in this case would require an averaging over a distribution of values for $r$.
Given the narrowness of the allowed $r$ values for the second solution
it seems highly unlikely that realistic averaging
would result in a solution satisfying
all the light element abundances (Meyer et al.\ 1991)
notwithstanding the problem with \li7.

Abundances of \li6, \be9, \b{10} and \b{11} are shown in
figures 3--6. With the exception of \be9
these are maximal abundances for all values of $R$, $f_v$.
Curves are given for $\eta_{10}=3.0$, 7.0, and 70.0.
Yields of these elements are again independent
of $r$ for $r\lsim100$ cm, indicating that the yields are unchanged from
those of homogeneous nucleosynthesis.  For $\eta_{10}=3$ this gives a \li6
abundance (number density relative to H) of roughly $3\times10^{-14}$, a
factor of 100 lower than the recent measurement of Smith, Lambert, \&
Nissen (1992).   Allowing for a higher \he4 abundance
and abandoning the \li7 constraints (that is, the $\eta_{10}\sim7$,
$r\sim10^4$ solution mentioned earlier) increases the yield by (at most)
a factor of 10. Of course if we now require the depletion of \li7,
\li6 will be severely depleted (Brown and Schramm 1988) and the discrepancy is
amplified.

The \be9 abundance (figure 4) is shown as maximal abundances for
$R=100,\,1000$, $f_v=1/64,\,1/8$ (solid curve) and as abundances for
$R=10^6$, $f_v=1/64$ (dashed curve). In this case, the effect of increasing $R$
was greatest.
For $r \lsim 100$ cm, the  \be9 abundance is $1\times10^{-18}$, four orders of
magnitude below the observations (Rebolo
et al.\ 1988; Ryan et al.\ 1992; Gilmore, Edvardsson, \&
Nissen 1991;).  Allowing
for $\eta_{10}\sim5$, $r\sim10^4$ cm raises this almost to $10^{14}$
(higher if we accept a density contrast of $R=10^6$),
however we regard this as an extremely unlikely situation.
Note that even though \be9 reaches a maximum at a few $\times 10^{-14}$ for
$\eta_{10} = 70$,
the other light elements are irreconcilably off from their measured abundances,
and this case is thus not viable.

\b{10} (figure 5) and \b{11}
(figure 6) have abundances of $\sim10^{-19}$ and $\sim10^{-17}$, respectively
7 and 5 orders of magnitude below observations (Duncan, Lambert, \&
Lemke 1992).  (The \b{10} abundance is always negligible compared to the
\b{11} abundance which is problematic if one wishes to show that the observed B
is primordial;
the two isotopes are observed with comparable abundances.)
Using $\eta_{10}\sim5$
and $r\sim10^4$ raises both abundance by less than two orders of
magnitude.  We emphasize that high abundances of \b{11} are produced only
for large $\eta$, and at a value of $r$ where the \be9 abundance is low.
Thus it is not possible to reconcile the large $r$ model with the
observed B to Be ratio, regardless of the problems with the light elements.

We also show maximum (solid) and minimum (dashed) yields of \he4 and \li7
in figures 7 and 8.  The possibility of a low \he4 abundance in inhomogeneous
models was also
investigated in Mathews, Schramm, \& Meyer (1993).

\vfill\eject

\section{4. Post-Processing}
We implemented the hydrodynamic-Thomson-drag dissipation effect by running the
inhomogeneous code to a mixing temperature $T_m = 20$ keV, the favored
value given
in Alcock et al.\ (1990).  We then homogenized the results and continued
to run  down to the usual final temperature
$T_f = 10^7 {\rm K} = 1.2$ keV.  While the effects of post-processing on
the lightest element (\D, \he3, \li7) depend rather strongly on the  input
parameters, we did find consistent results for Be and B.  The result is
a reduction in the yield of \be9 and \b{10}, and an increase in \b{11}.
While it is conceivable that the right set of parameters might bring this
model into agreement with the observations of the very light elements,
the increase in B combined with a decrease in Be is difficult to reconcile
with the observations (Duncan et al., 1992).  Consequently, we regard this
as an unlikely scenario.  Similar conclusions have been reached in the
recent parameter-free hydrodynamic calculations of Jedamzik \&
Fuller (1993).

\section{5. Conclusions}
With the accepted limits on the light element abundances (\he4, \h2, \he3,
\li7) the length scale of inhomogeneities at the epoch of primordial
nucleosynthesis is constrained to be $r\lsim100$ cm (at 100 MeV).
With this constraint, the abundances of the light elements, and of the
additional elements \li6, \be9, \b{10} and \b{11} are largely indistinguishable
from those of homogeneous nucleosynthesis.  In particular, the abundances
of LiBeB are lower (by several orders of magnitude) than the lowest of the
abundances seen recently in population II halo stars.  We conclude that
these elements must be produced by some process other than primordial
nucleosynthesis.

If we push the limits on the light element abundances to the extreme,
we find that while the abundances of LiBeB all increase, only \be9 is raised
significantly and still falls short of being able to explain any of
the recent observations.

\vfill\eject
\section{Acknowledgements}
This work was supported in part by the NSF, by the DOE and by NASA
at the University of Chicago,  by the DOE and by NASA theory grant
2381 at Fermilab, by DOE grant DE-AC02-83ER-40105 and by a Presidential
Young Investigator Award at the University of Minnesota.  Work performed
in part under the auspices of the U.S. Department of Energy at Lawrence
Livermore National Laboratory under contract number W-7405-ENG-48 and
Nuclear Theory grant SF-ENG-48.

\section{References}
\frenchspacing
\refitem Alcock, C.R., Dearborn, D.S., Fuller, G.M., Mathews, G.J., \&
        Meyer, B.S., 1990, ApJ, 64, 2607
\refitem Alcock, C., Fuller, G.M., \& Mathews, G.J. 1987, ApJ, 320, 439
\refitem Applegate, J.H., \& Hogan, C.J. 1985, Phys. Rev., D31, 3037;
        erratum in D34, 1938
\refitem Applegate, J.H., Hogan, C.J., \& Scherrer, R. 1987, Phys. Rev.,
        D35, 1151
\refitem Campbell, B.A., Ellis, J., \& Olive, K.A. 1990, Nucl. Phys. B345, 57
\refitem Crawford, M., \& Schramm D.N. 1982, Nature, 298, 538
\refitem Banerjee, B., \& Chitre, S.M. 1991, Phys. Lett., B258, 247
\refitem Boyd, R.N., \& Kajino, T. 1989, ApJ, 336, L55
\refitem Brown, L. \& Schramm, D.N. 1988, ApJ, 329,L103
\refitem Deliyannis, C. P., Demarque, P., \& Kawaler, S. D. 1990 ApJS, 73, 21
\refitem Duncan, D.K., Lambert, D.L., \& Lemke, M. 1992,
        ApJ, 401, 584
\refitem Fahri, E. \& Jaffe, R.L. 1984, Phys. Rev. D30, 2379
\refitem Fields, B., Schramm, D.N., \& Truran J. 1993, Fermilab preprint 92-114
\refitem Fuller, G.M., Mathews G.J., \& Alcock, C.R. 1988, Phys. Rev.,
        D37, 1380
\refitem Fowler, W.A., \& Hoyle, F. 1964, ApJS, 9, 201
\refitem Gilmore, G., Edvardsson, B., \& Nissen, P.E. 1991,
        ApJ, 378, 17
\refitem Gnedin, N.Yu., \& Ostriker, J.P., 1992, ApJ, 400, 1
\refitem Heckler A. \& Hogan, C.J. 1993, Phys. Rev. D, submitted
\refitem Hobbs, L.M., \& Duncan, D.K. 1987, ApJ, 317, 796
\refitem Hogan, C. 1983, Phys. Lett., B133,172
\refitem Iso, K., Kodama, H., \& Sato, K. 1986, Phys. Lett., B169, 337
\refitem Jedamzik, K., Fuller, G.M., \& Mathews, G.J. 1993a, ApJ, submitted
\refitem Jedamzik, K., Fuller, G.M., Mathews, G.J., \& Kajino, T.
	1993b, ApJ, in press
\refitem Jedamzik, K., \& Fuller, G.M., 1993, ApJ, in press.
\refitem Kajantie, K., Karkkainen, L., \& Rummukainen, K., 1990,
        Nucl.\ Phys., B333, 100.
\refitem Kapusta, J., \& Olive, K.A. 1988, Phys. Lett., B209, 295
\refitem Kurki-Suonio, H. 1988, Phys. Rev., D37, 2104
\refitem Kurki-Suonio, H., Aufderheide, M.B., Graziani, F., Mathews, G.J.,
        Banerjee, B., Chitre, S.M., \& Schramm, D.N. 1992, Phys. Lett. B,
        289, 211
\refitem Kurki-Suonio, H., \& Matzner, R.A. 1989, Phys. Rev. D,
        39, 1046
\refitem Kurki-Suonio, H., \& Matzner, R.A. 1990, Phys. Rev. D,
        42, 1047
\refitem Kurki-Suonio, H., Matzner, R.A., Olive, K.A., \&
        Schramm, D.N. 1990, ApJ, 353,406
\refitem Malaney, R.A., \& Fowler, W.A. 1988, ApJ, 333, 14
\refitem Malaney, R.A., \& Fowler, W.A. 1989, ApJ, 345, L5
\refitem Malaney, R.A., \& Mathews, G.J. 1993, Phys. Rep. (in press)
\refitem Mathews, G.J., Meyer, B.S., Alcock, C.R., \& Fuller, G.M. 1990,
        ApJ, 358, 36
\refitem Mathews, G.J., Schramm, D.N., \& Meyer, B.S., 1993, ApJ, 404, 476
\refitem Meyer, B.S., Alcock, C.R., Mathews, G.J., \& Fuller, G.M. 1991,
        Phys.\ Rev.\ D43, 1079
\refitem Olive, K. A., \& Schramm, D. N. 1992 Nature, 360,439
\refitem Particle Data Group, 1992, Phys.\ Rev.\ D45.
\refitem Prantzos, N., Cass\'{e}, M., \& Vangioni-Flam, E. 1993 ApJ, 403, 630
\refitem Rebolo, R., Molaro, P., Abia, C., \& Beckman, J.E. 1988,
        A\&A, 193, 193
\refitem Rebolo, R., Molaro, P., \& Beckman, J.E. 1988, A\&A, 192, 192
\refitem Ryan, S.G., Norris, J.E., Bessell, M.S., \& Deliyannis, C.P.
        1992, ApJ, 388, 184
\refitem Sale, K.E., Mathews, G.J. 1986, ApJ Lett., 309, L1
\refitem Skillman, E. et al. 1993, in {\it Texas/PASCOS 92: Relativistic
         Astrophysics and Particle Cosmology} eds. C.W. Akerlof \& M.A.
         Srednicki, (New York Academy of Sciences, New York) p.739
\refitem Smith, V.V., Lambert, D.L., \& Nissen, P.E. 1992, University of
        Texas preprint
\refitem Spite, F., \& Spite, M. 1982a Nature, 297, 483
\refitem Spite, F., \& Spite, M. 1982b A\&A, 115, 357
\refitem Spite, F., \& Spite, M. 1986 A\&A, 163, 140
\refitem Steigman, G., Fields, B., Olive, K.A., Schramm, D.N.,
\& Walker, T. 1993, University of Minnesota preprint UMN-TH-1123/93
\refitem Steigman, G., \& Walker, T.P. 1992, ApJ, 385, L13
\refitem Terasawa, N., \& Sato, K. 1989a, Phys. Rev., D39, 2893
\refitem Terasawa, N., \& Sato, K. 1989b, Prog. Theor. Phys., 81, 254
\refitem Terasawa, N., \& Sato, K. 1989c, Prog. Theor. Phys., 81, 1085
\refitem Thomas, D., Schramm, D.N., Olive, K.A., \& Fields, B.D. 1993,
        ApJ, 406, 569 (TSOF)
\refitem Wagoner, R.V. 1973, ApJ, 179,343
\refitem Walker, T.P., Steigman, G., Schramm, D.N., Olive, K.A., \&
        Fields, B. 1993, preprint (Fermilab-Pub-92/199-A) ApJ (in press)
\refitem Walker, T.P., Steigman, G., Schramm, D.N., Olive, K.A., \&
        Kang, H.-S. 1991, ApJ, 376, 51
\refitem Witten, E. 1984, Phys. Rev., D30, 272
\refitem Yang, J., Turner, M.S., Steigman, G., Schramm, D.N., \& Olive, K.A.
        1984, ApJ, 281,493
\nonfrenchspacing

\vfill\eject

%------------------ Tables -----------------------------------------------

\singlespacing
\centerline{Table 1}
\centerline{Reactions changed since TSOF}

\def\rule{\medskip{\hrule width \hsize}\medskip}
\def\drule{\medskip{\hrule width \hsize}\vskip.1em{\hrule width \hsize}
        \medskip}
\settabs\+\indent&XXXXXXXXXX&\cr % Sample line

\drule\+&Reaction&Rate\cr\rule

% \centerline{\it Beta decays}

% AS85
\+&\be{12}($\beta^-$)\b{12}&
$34.31 \pm 0.03$
\cr

% Lederer
\+&\b{13}($\beta^-$)\c{13}&
$40.00 \pm 0.40$
\cr

% Lederer
\+&\b{14}($\beta^-$)\c{14}&
$43.05 \pm 3.21$
\cr

% Lederer
\+&\he{8}($\beta^-$)\li{8}&
$5.68 \pm 0.09$
\cr

% AS83
\+&\li{11}($\beta^-$)\be{11}&
$79.7 \pm 0.9$
\cr

% AS86
\+&\b{15}($\beta^-$)\c{15}&
$63  \pm 6$
\cr

% AS86
\+&\c{17}($\beta^-$)\n{17}&
$3.43 \pm 0.29$
\cr

% AS87
\+&\c{18}($\beta^-$)\n{18}&
$10.5^{+2.4}_{-4.0}$
\cr

% Lederer
\+&\n{18}($\beta^-$)\o{18}&
$1.10 \pm 0.06$
\cr

% AS87
\+&\c{19}($\beta^-$)\n{19}&
$57.8$% (theory)
\cr

% AS87
\+&\n{19}($\beta^-$)\o{19}&
$3.3 \pm 1.7$
\cr

% AS87
\+&\c{20}($\beta^-$)\n{20}&
$74.5$% (theory)
\cr

% AS87
\+&\n{20}($\beta^-$)\o{20}&
$6.9^{+1.4}_{-2.1}$
\cr

% Endt90
\+&\o{21}($\beta^-$)\f{21}&
$0.202 \pm 0.006$
\cr

% Endt90
\+&\o{22}($\beta^-$)\f{22}&
$0.308 \pm 0.67$
\cr

% Endt90
\+&\f{22}($\beta^-$)\ne{22}&
$0.1639 \pm 0.0015$
\cr

% Endt90
\+&\f{23}($\beta^-$)\ne{23}&
$0.31 \pm 0.06$
\cr

% Endt90
\+&\f{24}($\beta^-$)\ne{24}&
$2.04 \pm 0.48$
\cr

%\centerline{\it Neutron Capture}

% AS87
\+&\o{18}(n,$\gamma$)\o{19}&
$21 \pm 1 + 7.3 \times 10^5 \,T_9^{-3/2} \exp (-1.846/T_9)+1.3\times10^5$
\cr

% Lederer
\+&\b{11}(n,$\gamma$)\b{12}&
$1.3 \times 10^5 \,T_9^{-3/2} \exp (-0.2112/T_9)
+ 4.0 \times 10^5 \,T_9^{-3/2} \exp (-4.53/T_9)$
\cr
\+&&$+1.3\times10^5$
\cr

%  Lederer
\+&\b{12}(n,$\gamma$)\b{13}&
$1.3 \times 10^6 \,T_9^{-3/2} \exp (-1.64/T_9)+1.3\times10^5$
\cr

% Lederer
\+&\b{13}(n,$\gamma$)\b{14}&
$2.8 \times 10^4 \,T_9^{-3/2} \exp (-4.02/T_9)+1.3\times10^5$
\cr

% AS87
\+&\o{19}(n,$\gamma$)\o{20}&
$4.6 \times 10^6 \,T_9^{-3/2} \exp (-0.186/T_9)
+ 4.6 \times 10^6 \,T_9^{-3/2} \exp (-1.74/T_9)$
\cr
\+&&$+1.3\times10^5$
\cr

% Endt90
\+&\f{21}(n,$\gamma$)\f{22}&
$1.8 \times 10^6 \,T_9^{-3/2} \exp (-4.18/T_9)+1.3\times10^5$
\cr

\+&\be{11}(n,$\gamma$)\be{12}&
$1.3 \times 10^5$
\cr

\+&\b{14}(n,$\gamma$)\b{15}&
$1.3 \times 10^5$
\cr

\+&\c{16}(n,$\gamma$)\c{17}&
$1.3 \times 10^5$
\cr

\+&\c{17}(n,$\gamma$)\c{18}&
$1.3 \times 10^5$
\cr

\+&\c{18}(n,$\gamma$)\c{19}&
$1.3 \times 10^5$
\cr

\+&\c{19}(n,$\gamma$)\c{20}&
$1.3 \times 10^5$
\cr

\+&\n{17}(n,$\gamma$)\n{18}&
$1.3 \times 10^5$
\cr

\+&\n{18}(n,$\gamma$)\n{19}&
$1.3 \times 10^5$
\cr

\+&\n{19}(n,$\gamma$)\n{20}&
$1.3 \times 10^5$
\cr

\+&\o{20}(n,$\gamma$)\o{21}&
$1.3 \times 10^5$
\cr

\+&\o{21}(n,$\gamma$)\o{22}&
$1.3 \times 10^5$
\cr

% He 8 +  p
% He8 (p,n) Li8
% He8 (p,\gamma) Li9
\+&\he{8}(p,n)\li8&
$0.2874\times10^{11} \,T_9^{-2/3} \exp (  -6.4847/ T_9^{1/3} )$
\cr
\+&\he{8}(p,$\gamma$)\li9&
ditto
\cr

\rule

\vfill\eject

%-------------------------------------------------------------------------

\centerline{Table 1 (contd.)}

\drule\+&Reaction&Rate\cr\rule

%  Li11 +  p
% Li11 (p,n) Be11
% Li11 (p,\alpha) He8
% Li11 (p,\gamma) Be12
\+&\li{11}(p,n)\be{11}&
$0.1149\times10^{12} \,T_9^{-2/3} \exp (  -8.5850/ T_9^{1/3} )$
\cr
\+&\li{11}(p,$\alpha$)\he{8}&
ditto
\cr
\+&\li{11}(p,$\gamma$)\be{12}&
ditto
\cr

%  Be12 +  p
% Be12 (p,n) B12
% Be12 (p,\alpha) Li9
% Be12 (p,\gamma) B13
\+&\be{12}(p,n)\b{12}&
$0.3247\times10^{12} \,T_9^{-2/3} \exp ( -10.4242/ T_9^{1/3} )$
\cr
\+&\be{12}(p,$\alpha$)\li{9}&
ditto
\cr
\+&\be{12}(p,$\gamma$)\b{13}&
ditto
\cr

%  B13 +  p
% B13 (p,n) C13
% B13 (p,\alpha) Be10
% B13 (p,\gamma) C14
\+&\b{13}(p,n)\c{13}&
$0.7917\times10^{12} \,T_9^{-2/3} \exp ( -12.1202/ T_9^{1/3} )$
\cr
\+&\b{13}(p,$\alpha$)\be{10}&
ditto
\cr
\+&\b{13}(p,$\gamma$)\c{14}&
ditto
\cr

%  B14 +  p
% B14 (p,n) C14
% B14 (p,\alpha) Be11
% B14 (p,\gamma) C15
\+&\b{14}(p,n)\c{14}&
$0.8355\times10^{12} \,T_9^{-2/3} \exp ( -12.1408/ T_9^{1/3} )$
\cr
\+&\b{14}(p,$\alpha$)\be{11}&
ditto
\cr
\+&\b{14}(p,$\gamma$)\c{15}&
ditto
\cr

%  B15 +  p
% B15 (p,n) C15
% B15 (p,\alpha) Be12
% B15 (p,\gamma) C16
\+&\b{15}(p,n)\c{15}&
$0.8788\times10^{12} \,T_9^{-2/3} \exp ( -12.15887/ T_9^{1/3})$
\cr
\+&\b{15}(p,$\alpha$)\be{12}&
ditto
\cr
\+&\b{15}(p,$\gamma$)\c{16}&
ditto
\cr

%  C15 +  p
% C15 (p,n) N15
% C15 (p,\alpha) B12
% C15 (p,\gamma) N16
\+&\c{15}(p,n)\n{15}&
$0.1850\times10^{13} \,T_9^{-2/3} \exp ( -13.73032/ T_9^{1/3})$
\cr
\+&\c{15}(p,$\alpha$)\b{12}&
ditto
\cr
\+&\c{15}(p,$\gamma$)\n{16}&
ditto
\cr

%  C16 +  p
% C16 (p,n) N16
% C16 (p,\alpha) B13
% C16 (p,\gamma) N14
%\noindent{\it there might be a resonant contribution to this (next) one;
%I'm not sure if its going to matter}
\+&\c{16}(p,n)\n{16}&
$0.1950\times10^{13} \,T_9^{-2/3} \exp ( -13.74825/ T_9^{1/3})$
\cr
\+&\c{16}(p,$\alpha$)\b{13}&
ditto
\cr
\+&\c{16}(p,$\gamma$)\n{17}&
ditto
\cr

%  C17 +  p
% C17 (p,n) N17
% C17 (p,\alpha) B14
% C17 (p,\gamma) N18
\+&\c{17}(p,n)\n{17}&
$0.2049\times10^{13} \,T_9^{-2/3} \exp ( -13.76414/ T_9^{1/3})$
\cr
\+&\c{17}(p,$\alpha$)\b{14}&
ditto
\cr
\+&\c{17}(p,$\gamma$)\n{18}&
ditto
\cr

%  C18 +  p
% C18 (p,n) N18
% C18 (p,\alpha) B15
% C18 (p,\gamma) N19
\+&\c{18}(p,n)\n{18}&
$0.2147\times10^{13} \,T_9^{-2/3} \exp ( -13.77833/ T_9^{1/3})$
\cr
\+&\c{18}(p,$\alpha$)\b{15}&
ditto
\cr
\+&\c{18}(p,$\gamma$)\n{19}&
ditto
\cr

%  C19 +  p
% C19 (p,n) N19
% C19 (p,\gamma) N20
\+&\c{19}(p,n)\n{19}&
$0.2246\times10^{13} \,T_9^{-2/3} \exp ( -13.79108/ T_9^{1/3})$
\cr
\+&\c{19}(p,$\gamma$)\n{20}&
ditto
\cr

%  C20 +  p
% C20 (p,n) N20
% C20 (p,\gamma) N21
\+&\c{20}(p,n)\n{20}&
$0.2343\times10^{13} \,T_9^{-2/3} \exp ( -13.80259/ T_9^{1/3})$
\cr
\+&\c{20}(p,$\gamma$)\n{21}&
ditto
\cr

%  N17 +  p
% N17 (p,n) O18
% N17 (p,\alpha) C14
% N17 (p,\gamma) O18
\+&\n{17}(p,n)\o{17}&
$0.4024\times10^{13} \,T_9^{-2/3} \exp ( -15.25388/ T_9^{1/3})$
\cr
\+&\n{17}(p,$\alpha$)\c{14}&
ditto
\cr
\+&\n{17}(p,$\gamma$)\o{18}&
ditto
\cr

%  N18 +  p
% N18 (p,n) O18
% N18 (p,\alpha) C15
% N18 (p,\gamma) O19
\+&\n{18}(p,n)\o{18}&
$0.4235\times10^{13} \,T_9^{-2/3} \exp ( -15.26960/ T_9^{1/3})$
\cr
\+&\n{18}(p,$\alpha$)\c{15}&
ditto
\cr
\+&\n{18}(p,$\gamma$)\o{19}&
ditto
\cr

%  N19 +  p
% N19 (p,n) O19
% N19 (p,\gamma) O20
\+&\n{19}(p,n)\o{19}&
$0.4445\times10^{13} \,T_9^{-2/3} \exp ( -15.28373/ T_9^{1/3})$
\cr
\+&\n{19}(p,$\gamma$)\o{20}&
ditto
\cr

\rule

\vfill\eject

%-------------------------------------------------------------------------

\centerline{Table 1 (contd.)}

\drule\+&Reaction&Rate\cr\rule

%  N19 + He
% N19 (\alpha,p) O22
\+&\n{19}($\alpha$,p)\o{22}&
$0.5537\times10^{18} \,T_9^{-2/3} \exp ( -36.75954/ T_9^{1/3})$
\cr

%  N20 +  p
% N20 (p,n) O20
% N20 (p,\alpha) C17
% N20 (p,\gamma) O21
\+&\n{20}(p,n)\o{20}&
$0.4656\times10^{13} \,T_9^{-2/3} \exp ( -15.29649/ T_9^{1/3})$
\cr
\+&\n{20}(p,$\alpha$)\c{17}&
ditto
\cr
\+&\n{20}(p,$\gamma$)\o{21}&
ditto
\cr

%  O20 +  p
% O20 (p,n) F20
% O20 (p,\gamma) F21
\+&\o{20}(p,n)\f{20}&
$0.8705\times10^{13} \,T_9^{-2/3} \exp ( -16.72064/ T_9^{1/3})$
\cr
\+&\o{20}(p,$\gamma$)\f{21}&
ditto
\cr

%  O21 +  p
% O21 (p,n) F21
% O21 (p,\gamma) F20
\+&\o{21}(p,n)\f{21}&
$0.9127\times10^{13} \,T_9^{-2/3} \exp ( -16.73330/ T_9^{1/3})$
\cr
\+&\o{21}(p,$\gamma$)\f{22}&
ditto
\cr

%  O22 +  p
% O22 (p,n) F22
\+&\o{22}(p,n)\f{22}&
$0.9552\times10^{13} \,T_9^{-2/3} \exp ( -16.74484/ T_9^{1/3})$
\cr

%  F22 +  p
% F22 (p,\alpha) O19
\+&\f{22}(p,$\alpha$)\o{19}&
$0.1712\times10^{14} \,T_9^{-2/3} \exp ( -18.11268/ T_9^{1/3})$
\cr

\+&\b{11}($\alpha,\gamma$)\n{15}&
$0.4314\times10^{16}  \,T_9^{-2/3} \exp ( -28.22994/ T_9^{1/3})$
\cr

\+&\li9($\alpha$,n)\b{12}&
$0.6221\times10^{14}  \,T_9^{-2/3} \exp ( -19.70047/ T_9^{1/3})$
\cr

\+&\b{10}($\alpha,\gamma$)\n{14}&
$0.3251\times10^{16}  \,T_9^{-2/3} \exp ( -27.98338/ T_9^{1/3})$
\cr

\+&\b8($\alpha,\gamma$)\n{12}&
$0.1662\times10^{16}  \,T_9^{-2/3} \exp ( -27.34717/ T_9^{1/3})$
\cr

\+&\n{17}($\alpha$,p)\o{20}&
$0.3805\times10^{18}  \,T_9^{-2/3} \exp ( -36.51220/ T_9^{1/3})$
\cr

\+&\c{16}($\alpha,\gamma$)\o{20}&
$0.6796\times10^{17}  \,T_9^{-2/3} \exp ( -32.81660/ T_9^{1/3})$
\cr

\rule
\doublespacing

%------------------ Figure Captions --------------------------------------
\vfill\eject
\section{Figure Captions}

\item{1} The nuclear reaction network used in all calculations.
\item{2a} Limits on $r$ and $\eta$ due
        to the light element abundances, for $R=100$.  Curves show
        the most generous limits for $f_v=1/8$ and $f_v=1/64$, and
        represent the following abundances:  \h2/H $=1.8\times10^{-5}$,
        (\h2+\he3)/H $=1.0\times10^{-4}$, \li7/H $=1.4\times10^{-10}$,
        $Y_p=0.22,0.24$.  The dashed curve is for $Y_p=0.245$.
        The hatched area shows the region allowed by the light element
        abundances.
\item{2b} As figure 2, for $R=1000$.
\item{2c} As figure 2, but for $R=10^6$, $f_v=1/64$.
\item{3} \li6 abundance as a function of $r$ for $\eta_{10} = 3,7,70$
        ($\eta_{10}=70$ is included only for illustrative purposes, as
        the light element abundances can never all be fit in this case).
        The curves represent the most generous abundances for all values
        of $R$, $f_v$.  The hatched line shows the upper limit on $r$.
\item{4} As figure 3, with \be9 abundances, except that the solid curves
        represent the maximum yields for all of $R=100,1000$, $f_v=1/8,1/64$
        and the dashed curves represent yields for $R=10^6$, $f_v=1/64$.
\item{5} As figure 3, with \b{10} abundances.
\item{6} As figure 3, with \b{11} abundances.
\item{7} As figure 3, with \he4 abundances.  In addition, the dashed line
        shows the lowest yield for all $R, f_v$.
\item{8} As figure 7, with \li7 abundances.

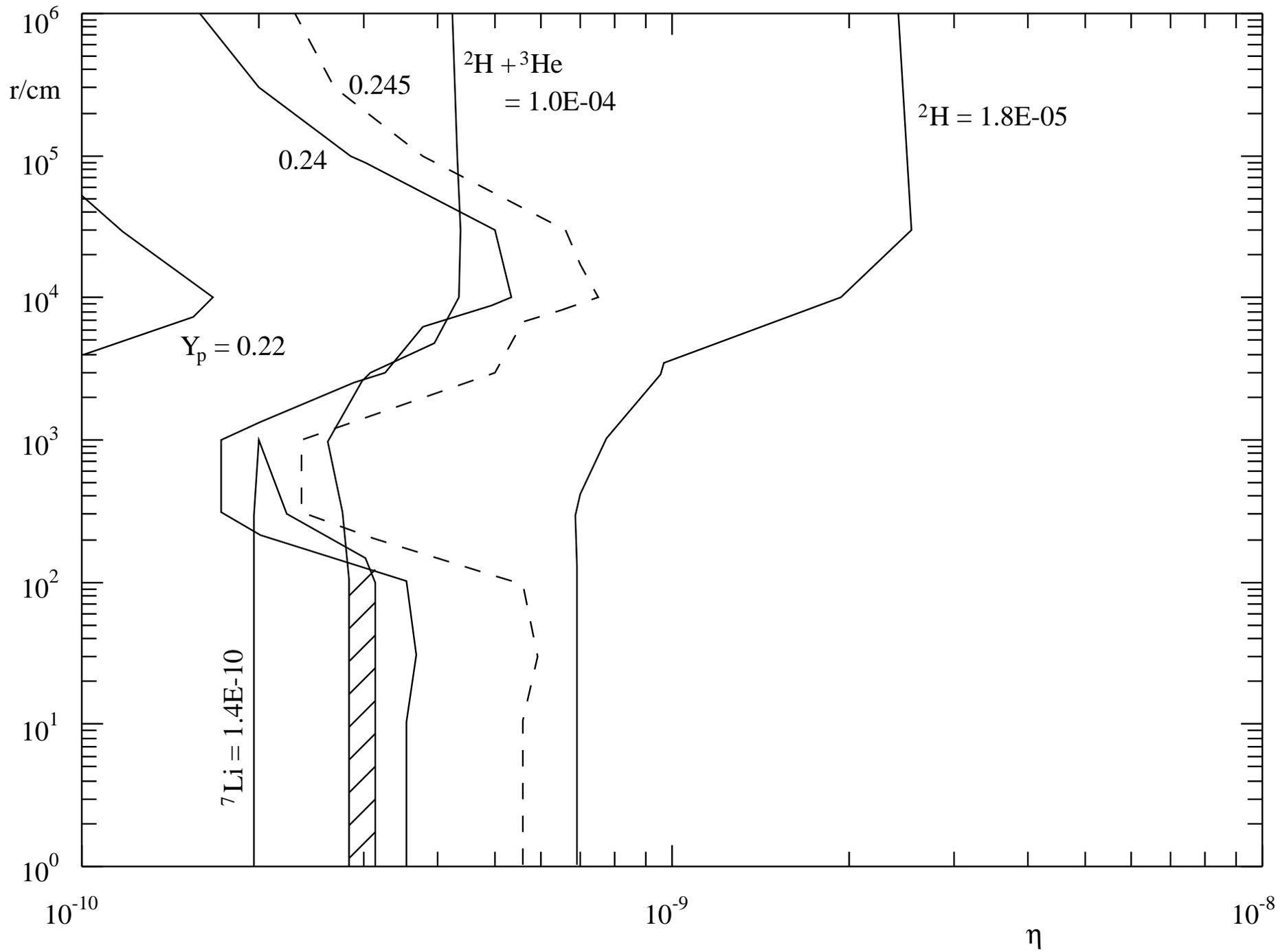
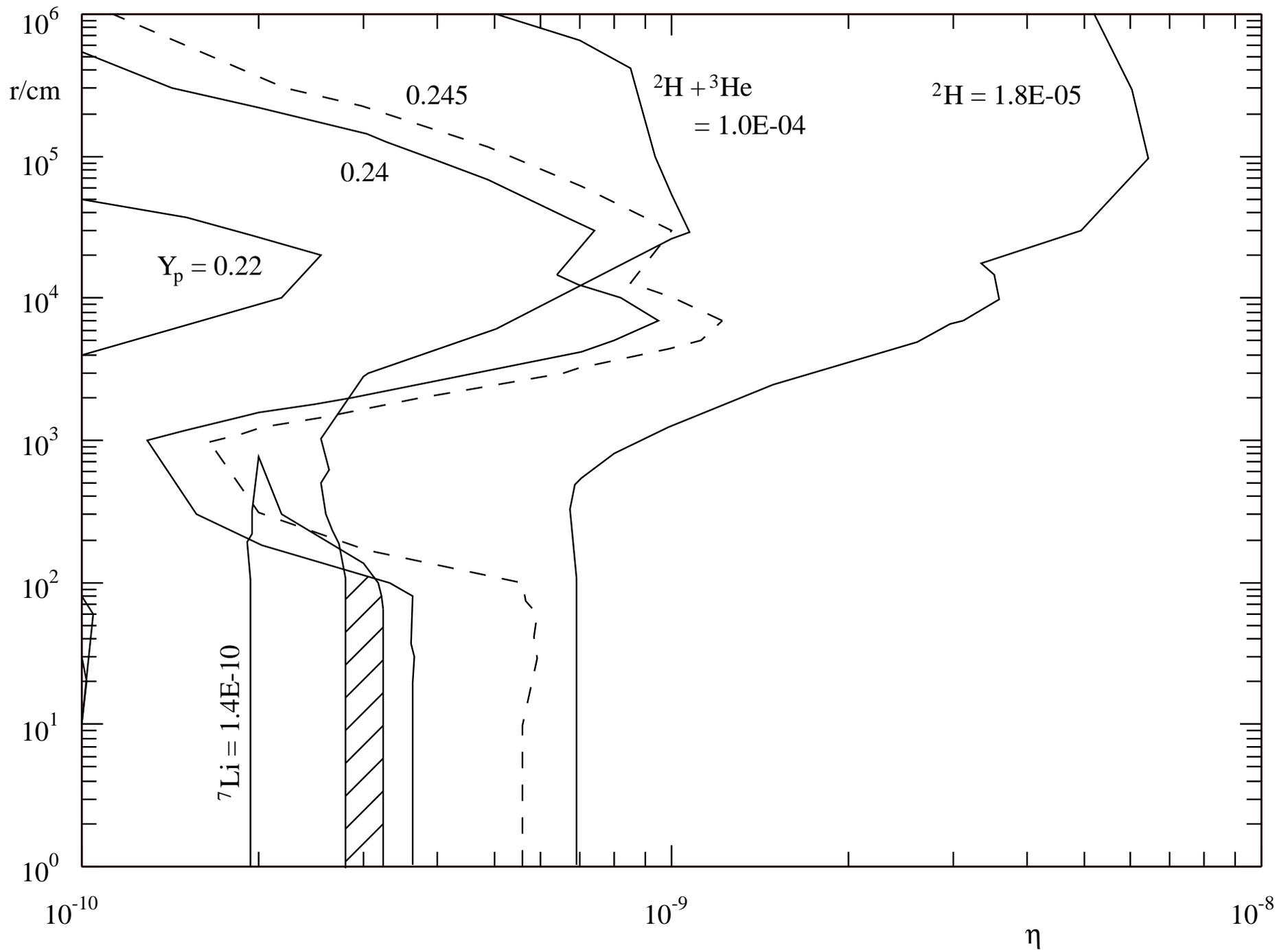
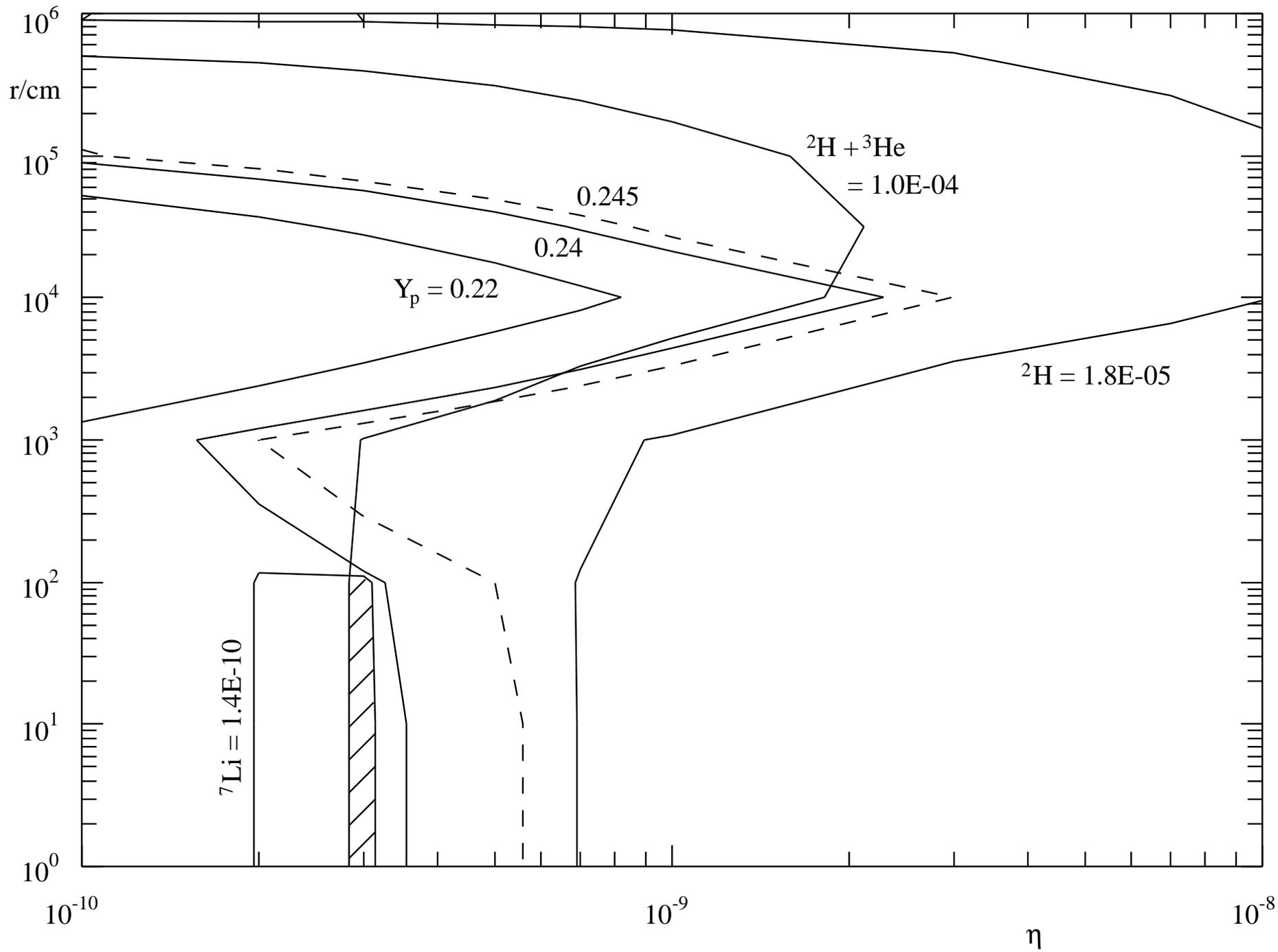
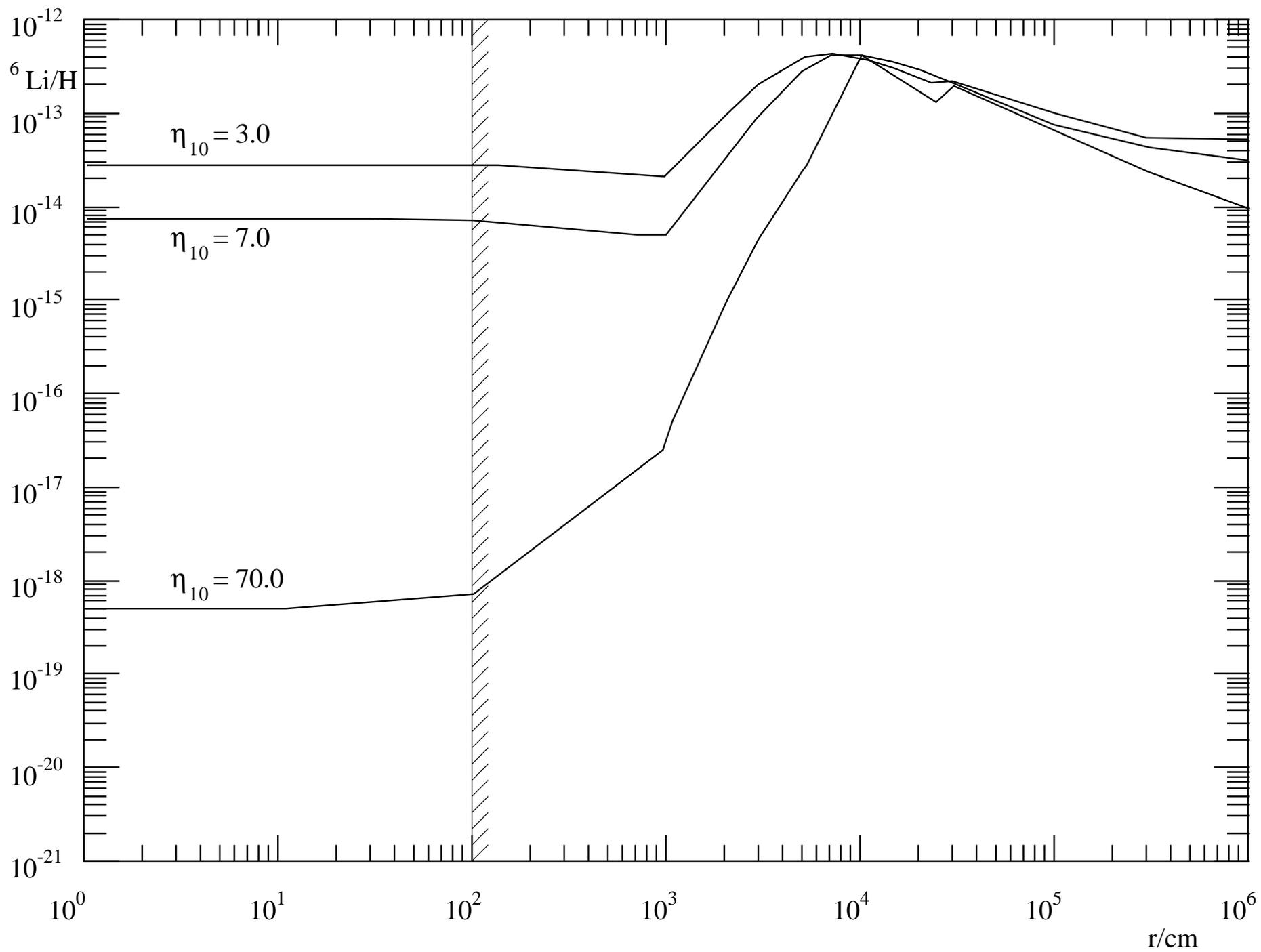
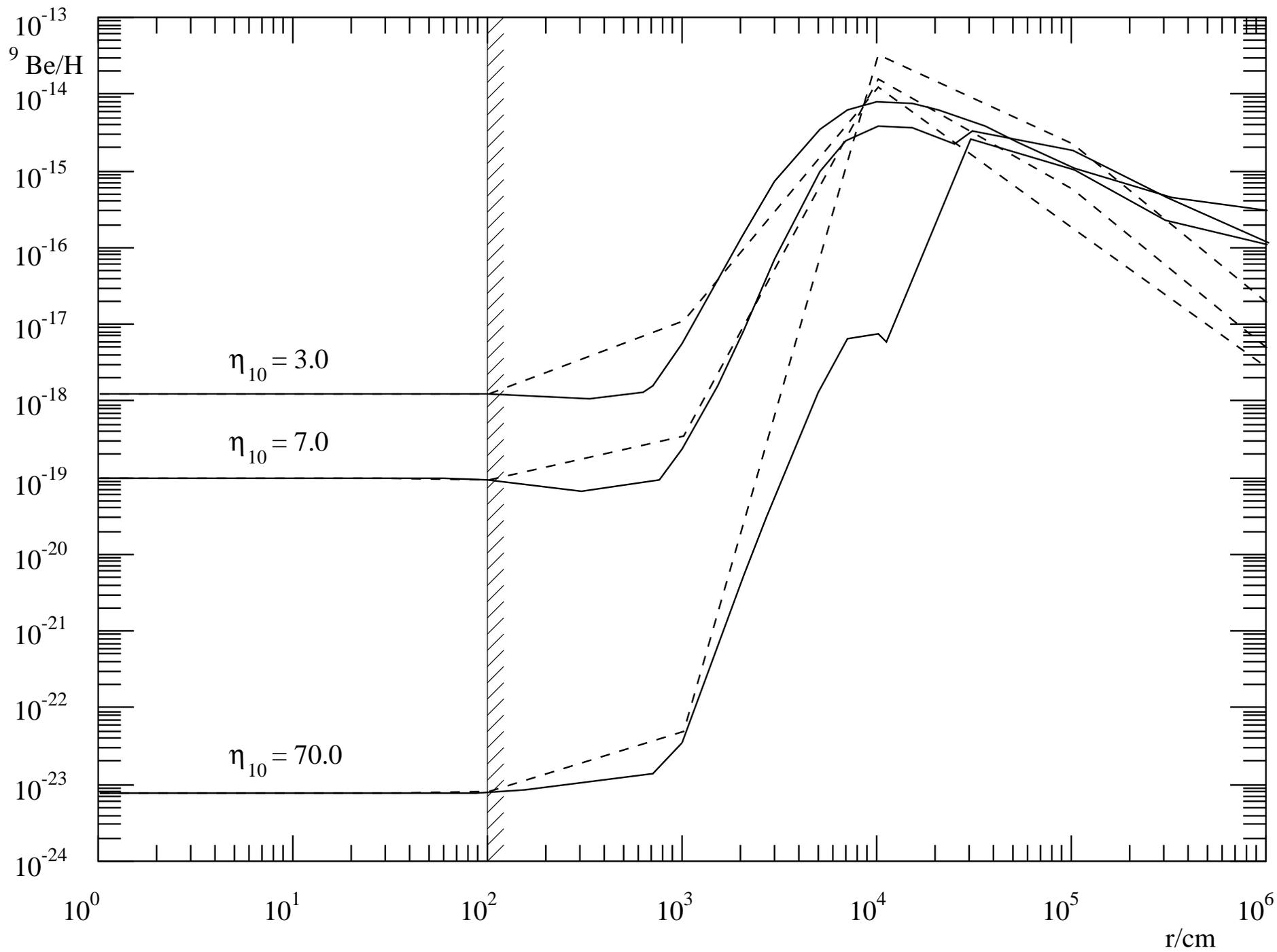
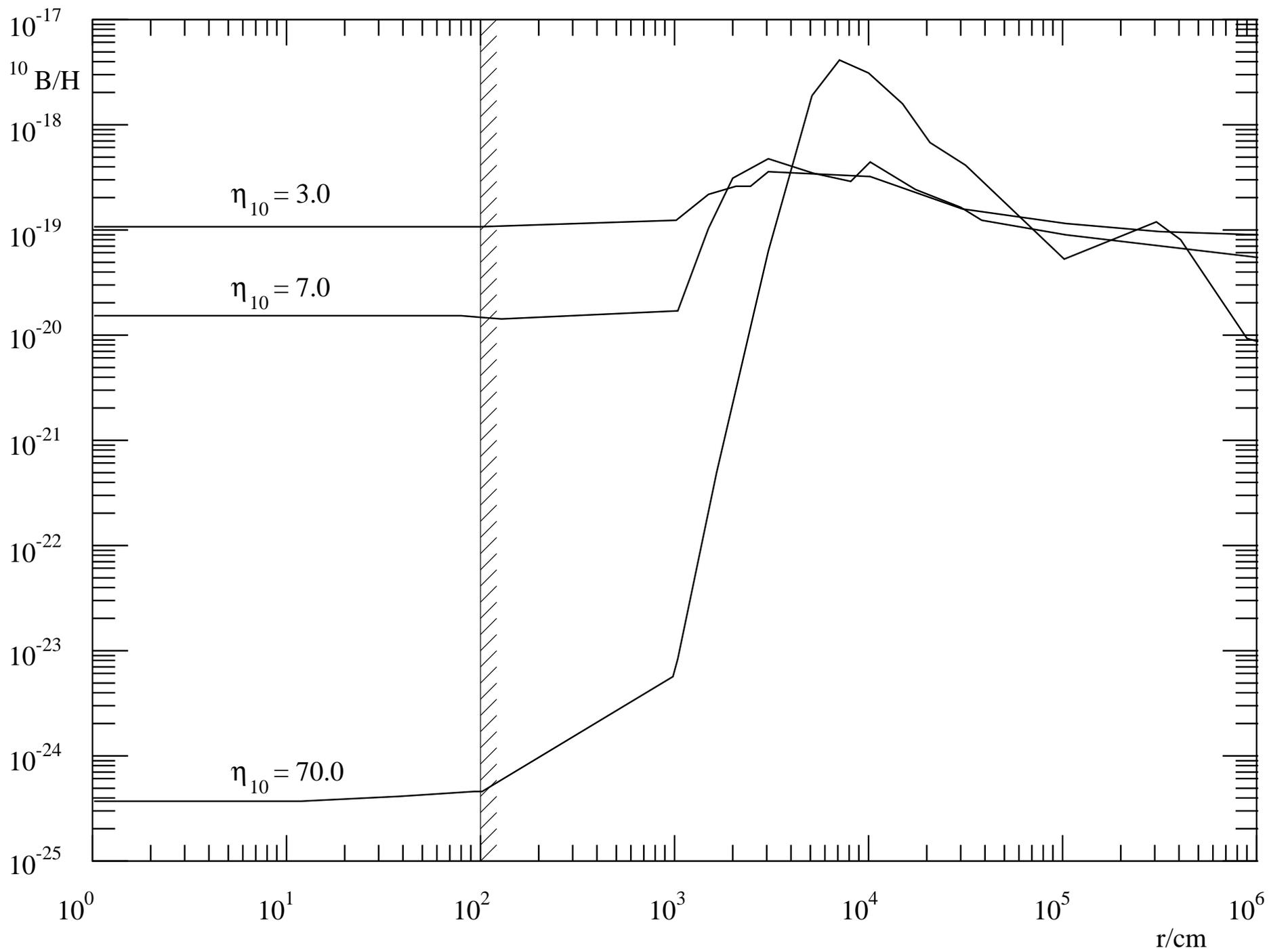
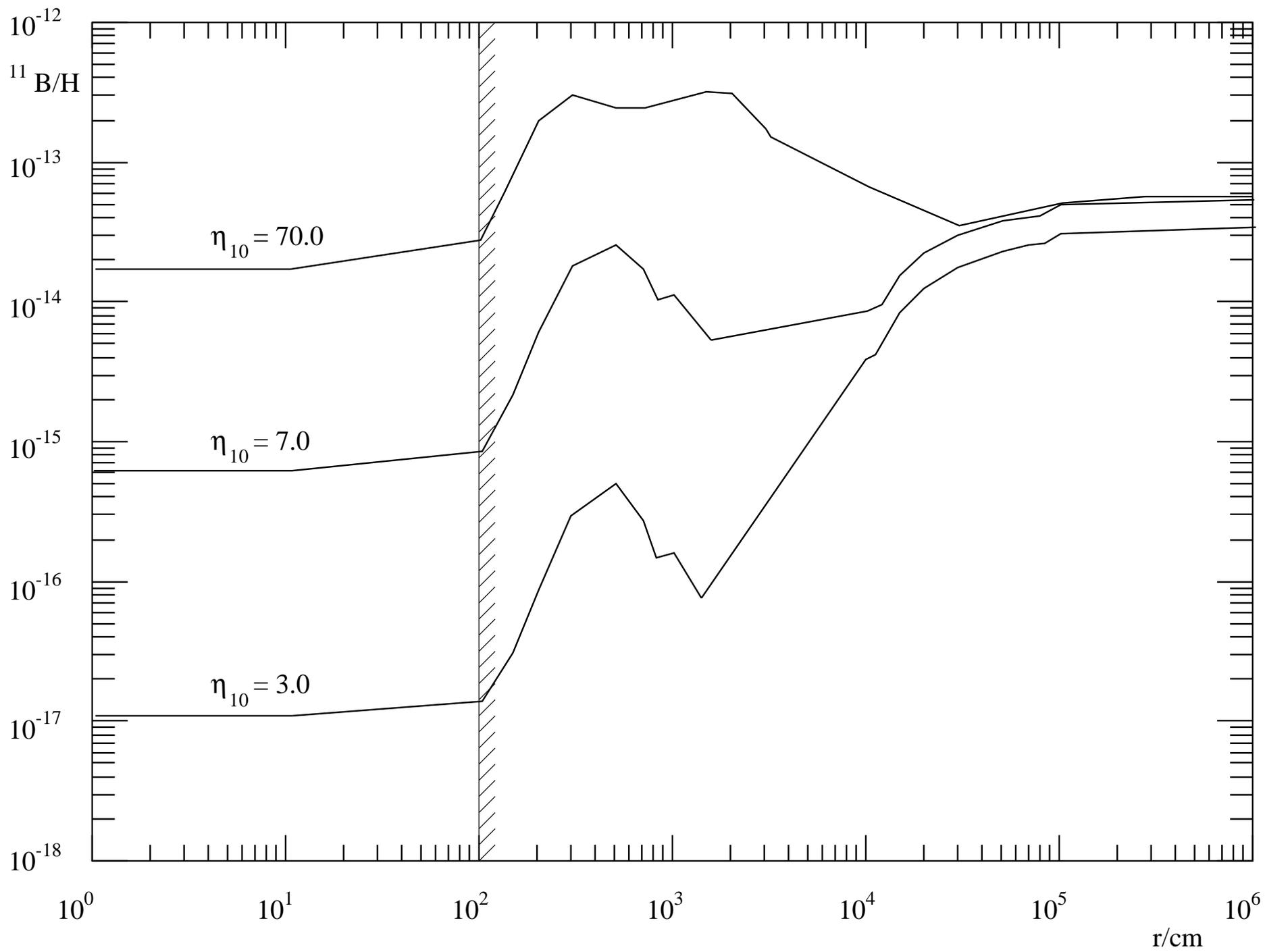
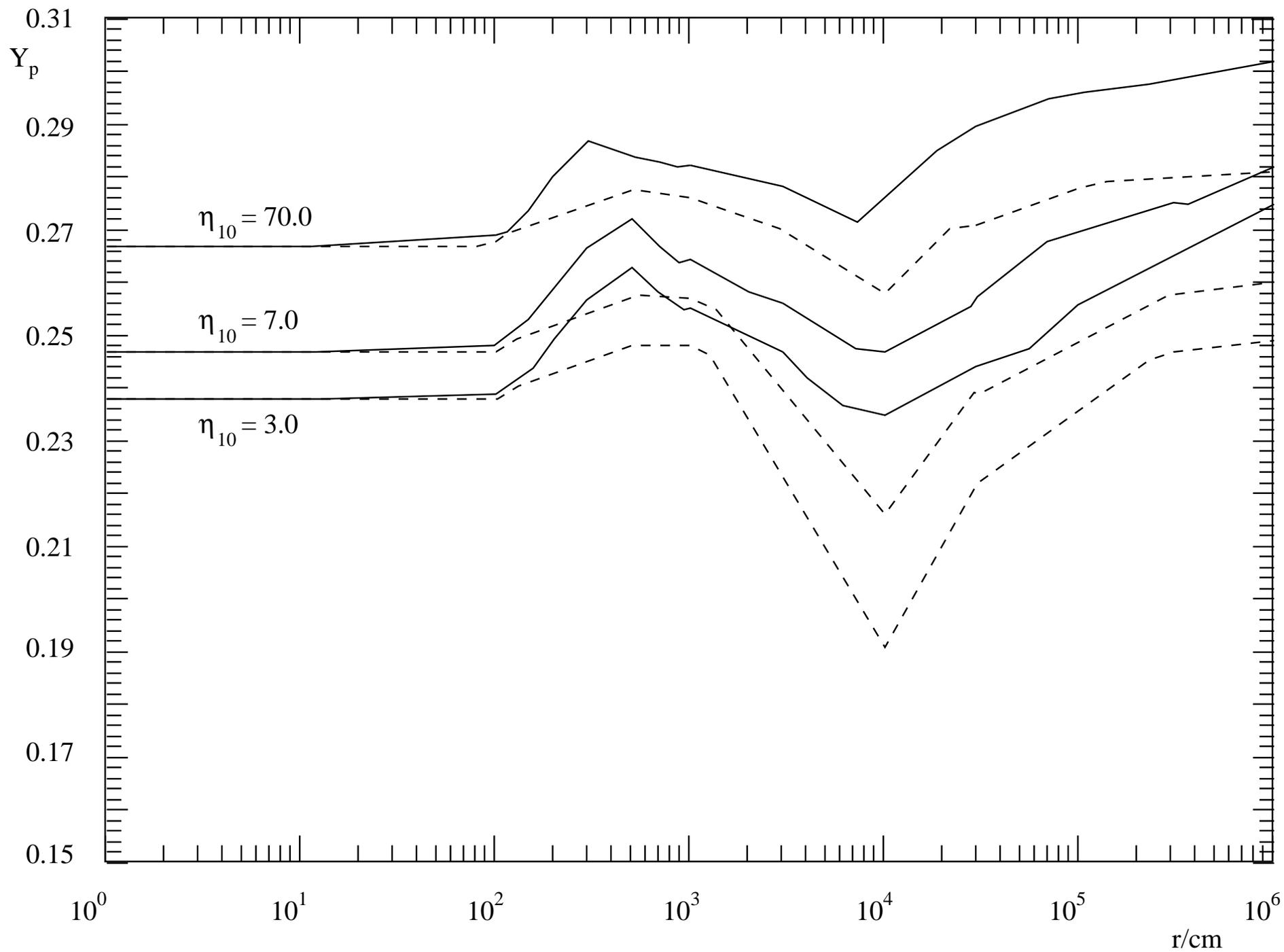
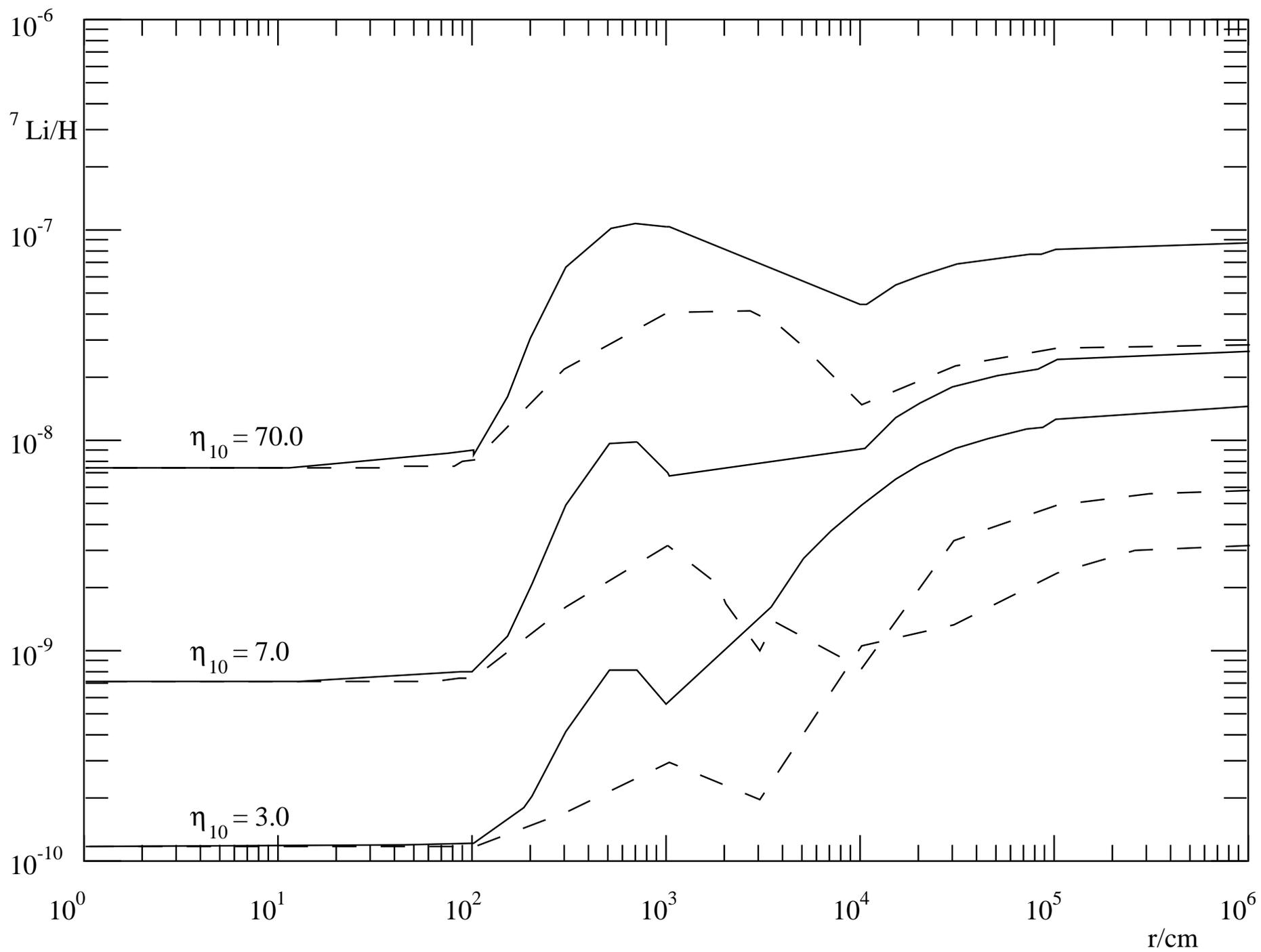
\bye